\documentstyle[11pt,epsfig]{article}
\author{M. Neek-Amal$^1$~\footnote{Corresponding author: mehdi.neekamal@gmail.com},  N. Abedpour$^2$ , S. N. Rasuli$^3$,
A. Naji$^{3}$ and M. R. Ejtehadi$^{2}$\\  {\small $^1$ Department of Physics, Shahid Rajaei University, Lavizan, Tehran 16788, Iran }\\
{\small$^2$  Department of Physics, Sharif University of Technology,
Tehran 11155-9161, Iran }\\{\small $^3$School of Physics, Institute
for Research in Fundamental Sciences, Tehran 19395-5531, Iran }\\}
\title{\bf Diffusive motion of $C_{60}$ on a graphene sheet}
\begin{document}
\maketitle \maketitle \vspace{0.5cm}
\begin{abstract}
The motion of a $C_{60}$ molecule over a graphene sheet at finite
temperature is investigated both theoretically and computationally.
We show that a graphene sheet generates a van der Waals laterally
periodic potential, which directly influences the motion of external
objects in its proximity. The translational motion of a $C_{60}$
molecule near a graphene sheet is found to be diffusive in the
lateral directions. While, in the perpendicular direction, the
motion may be described as diffusion in an effective harmonic
potential which is determined from the distribution function of the
position of the $C_{60}$ molecule. We also examine the rotational
diffusion of $C_{60}$ and show that its motion over the graphene
sheet is not a rolling motion.
\end{abstract}


\pagebreak
\section{Introduction}
Various properties of graphene as a new two dimensional material
have been studied both experimentally and
theoretically~\cite{novoselov,geim}. A recent experimental research
based on TEM visualization studies the images and dynamics of light
atoms deposited  on a single-layer graphene sheet~\cite{nature2008}.
On the other hand diffusion process and crystallization of atoms and
light molecules on various surfaces have also been subject of
research for many years due to both their theoretical importance as
well as their technological applications. The theoretical studies of
the motion of molecular scale objects on the various surface is also
applicable to the motion of nanoscale object's over
nanoelectromechanical surfaces, i.e., graphene. These studies play
an important role in designing graphene based nanosensors. Diffusive
motion of inclusions ({\em e.g.} macromolecules) over a rough
membrane is another related subject, which has received a lot of
attention in recent years~\cite{physrev2007,EPL2005,alinaji}.
Solidification of $C_{60}$ molecules over various substrates
particulary graphene is another subject of interest. The van der
Waals epitaxy of a solid $C_{60}$ over graphene sheet has been done
in recent experiments \cite{hashimoto}. Also various dynamical
properties of the spinning motion of $C_{60}$ on the Au(1 1 1)
surface have been studied by Teobaldi \emph{et al.} by means of
molecular dynamics (MD) simulations~\cite{small2007}.

The choice of van der Waals parameters of $C_{60}$--graphene
interaction is an important issue for performing any  molecular
dynamics simulations. The $C_{60}$--carbon nanotubes interaction can
be obtained both experimentally and theoretically~\cite{Hendrick}.
Several van der Waals parameters for physical adsorption of $C_{60}$
on graphite and other substrates were formulated using a continuum
rigid body model for $C_{60}$ and a continuum dielectric media for
graphite by Girad \emph{et al}~\cite{girad}. The charge transfer
from graphene to the $C_{60}$ molecule is an open question which may
be tackled using modern density functional theories. For alkali
metals on the graphite there have been some calculations in order to
estimate transferred charges~\cite{jcondmat}.

Two main approaches for investigating the diffusive motion of an
external inclusion on the membranes surface can be introduced. The
first one is based on the continuum description for the membrane
structure and introduces an effective Hamiltonian which can be used
to study curvature-coupled
diffusion~\cite{EPL2005,physrev2007,alinaji}. The elastic aspects of
the membrane  thus play a key role in this approach. The second
approach is an atomistic one which considers the details of the
membrane and inclusion structures as well as atomic interactions on
different levels of coarse graining~\cite{Skim,farago}. The main
goal in these studies is to characterize the inclusion's motion.

The diffusive motion of particles may be modeled generally via a
Langevin equation.
Lacasta \emph{et al} have modeled general two dimensional solid
surfaces using a two dimensional potential both with a deterministic
periodic potential and a random one~\cite{subdiffusion}. Different
values for dissipation parameters generate different trajectories
for a point-like particle over such potentials, which thus lead to
different dynamical behaviors ranging from subdiffusive to
superdiffusive motion~\cite{subdiffusion}.

In a previous study we showed that the alkali and transition metals
distribute over graphene and bind to its surface via a Lennard-Jones
potential and construct atomic nanoclusters~\cite{neek2009}. For
metallic nanoclusters,  no diffusive motion was found in low
temperature. It is also found that potassium atoms, in low
temperature, construct a particular phase on the graphene sheet as
well as on the graphite~\cite{jcondmat}. To our knowledge there have
been only a few number of studies on the interaction between
graphene and inclusions that address the pattern of inclusion
motion~\cite{physrev2007,EPL2005,alinaji}.

In this paper we study the dynamics of a single molecule,
specifically chosen as the $C_{60}$ molecule,
 on a graphene surface.
We show that a graphene sheet creates a periodic van der Waals
potential in its surrounding space and that there are some simple
criteria to determine when a molecule may diffuse through the
potential wells generated by the graphene (as assisted by thermal
fluctuations) or be likely to be trapped in the potential wells near
the surface. We introduce a laterally-averaged effective potential
for the graphene sheet from the distribution function of $C_{60}$
near the surface, and show that this potential may be approximated
best by a harmonic potential in the direction normal to the sheet.
We also introduce an effective friction coefficient for the
diffusive motion of $C_{60}$ over graphene. Finally, we show that
the motion of $C_{60}$ over
graphene is not a rolling motion 
and also the variation of the perpendicular component of angular
velocity of $C_{60}$ is greater than its component parallel to the
graphene sheet.

\section{Methods}

We employ classical Molecular Dynamics (MD) algorithm to simulate
the $C_{60}$-graphene system. The graphene is modeled as a
square-like sheet of area 1883 nm$^2$ constructed of 72000 carbon
atoms. The temperature was kept constant (300K) in our simulations
by employing a Nos\'e-Hoover thermostat. For the covalent bounds
between the nearest neighbor atoms of the graphene sheet (and for
chemically bonded atoms in $C_{60}$ molecule) we have used Brenner's
potential \cite{brenner}. For the interaction between each atom of
graphene with each of $C_{60}$'s atoms, we have used the
Lennard-Jones potential (LJ),
$$U_{LJ}=4\varepsilon\{(\sigma/r)^{12}-(\sigma/r)^{6}\}$$ with a
typical values $\sigma=3.4$~\AA\, and
$\varepsilon=2.4$~meV~\cite{rafiitabar}. A graphene sheet was
initially positioned, on average, at $z=0$ plane, and the centers of
the $C_{60}$ molecule were placed above it at $z=7$~\AA. The
temperature-dependent molecular dynamics simulations run for up to
two nanoseconds.

\begin{figure}[ht]
\begin{center}
\includegraphics[width=0.3\linewidth]{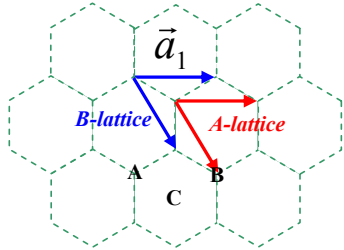}
\caption{(Color online) Two common triangular sublattices of
graphene sheet in the $x-y$ plane. \label{figAB}}
\end{center}
\end{figure}

\begin{figure}[ht]
\begin{center}
\includegraphics[width=0.45\linewidth]{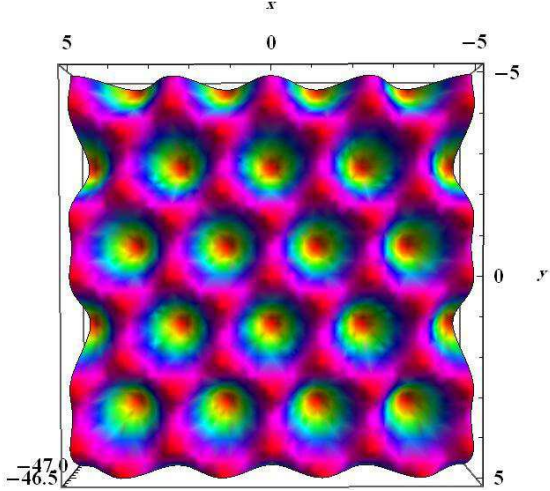}
\caption{(Color online) Periodic two dimensional potential,
$\tilde{V}(x,y)$, created by the graphene sheet. Here $x$ and $y$
refer to spatial position at the height $z_0$=3.5\AA\, above the
graphene sheet. This figure shows the top view of the potential.
\label{fig2Dp}}
\end{center}
\end{figure}
\begin{figure}[ht]
\begin{center}
\includegraphics[width=0.6\linewidth]{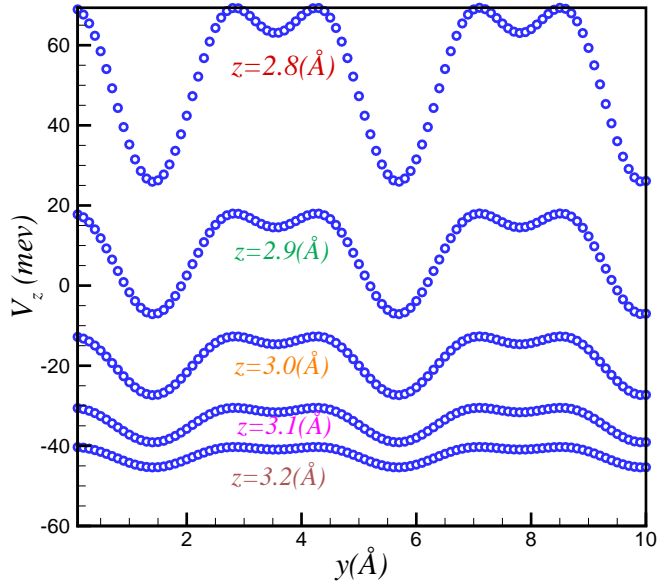}
\caption{(Color online) Periodic potential energy between flat
graphene sheet and a single carbon atom at $x$=0 as a function of y
for several height values.
 \label{vy}}
\end{center}
\end{figure}

\begin{figure}[ht]
\begin{center}
\includegraphics[width=0.6\linewidth]{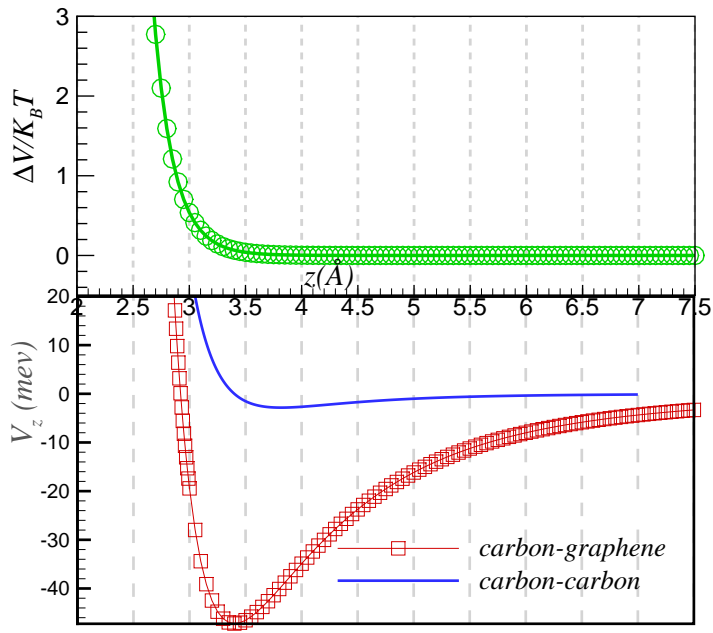}
 \caption{(Color online) Top: Difference between maximum and minimum of
two-dimensional potential (averaged over Wigner-Seitz primitive cell
of both A-lattice and B-lattice) created by a flat graphene sheet
(see Fig. \ref{fig2Dp}) as a function of normal distance $z$ as
experienced by a single carbon atom. Bottom: Total potential energy
between a flat graphene sheet and a single carbon atom has been
depicted. The solid blue curve is a simple LJ potential between two
carbon atoms. In both panels, to eliminate the dependence on the x
and y variables, we averaged over the first Brillouin zone of both A-lattice and B-lattice. The data have
small ($10^{-5}$) error bars which are not shown.\label{vz} }
\end{center}
\end{figure}

\begin{figure}[ht]
\begin{center}
\includegraphics[width=0.45\linewidth]{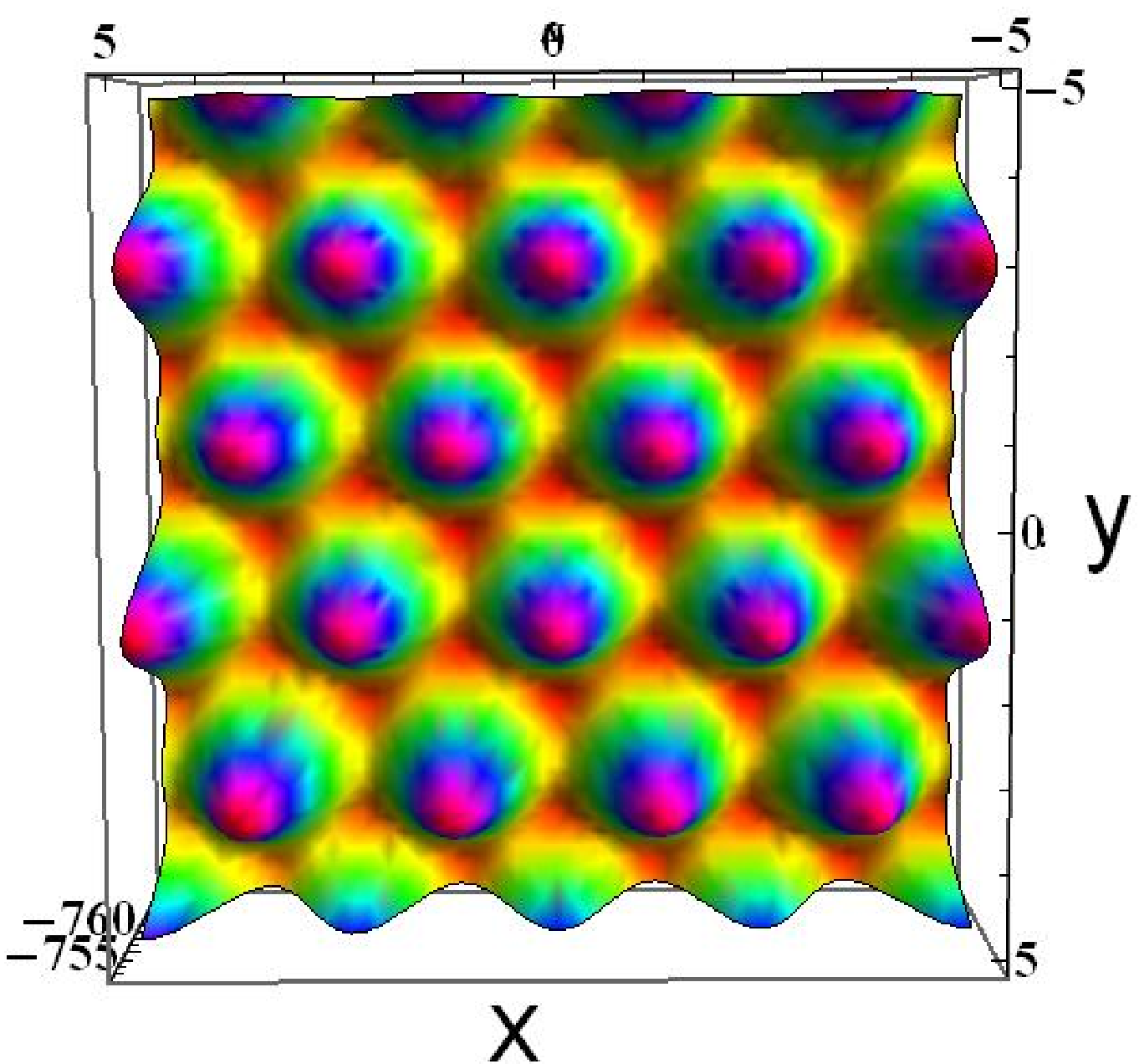}
\includegraphics[width=0.45\linewidth]{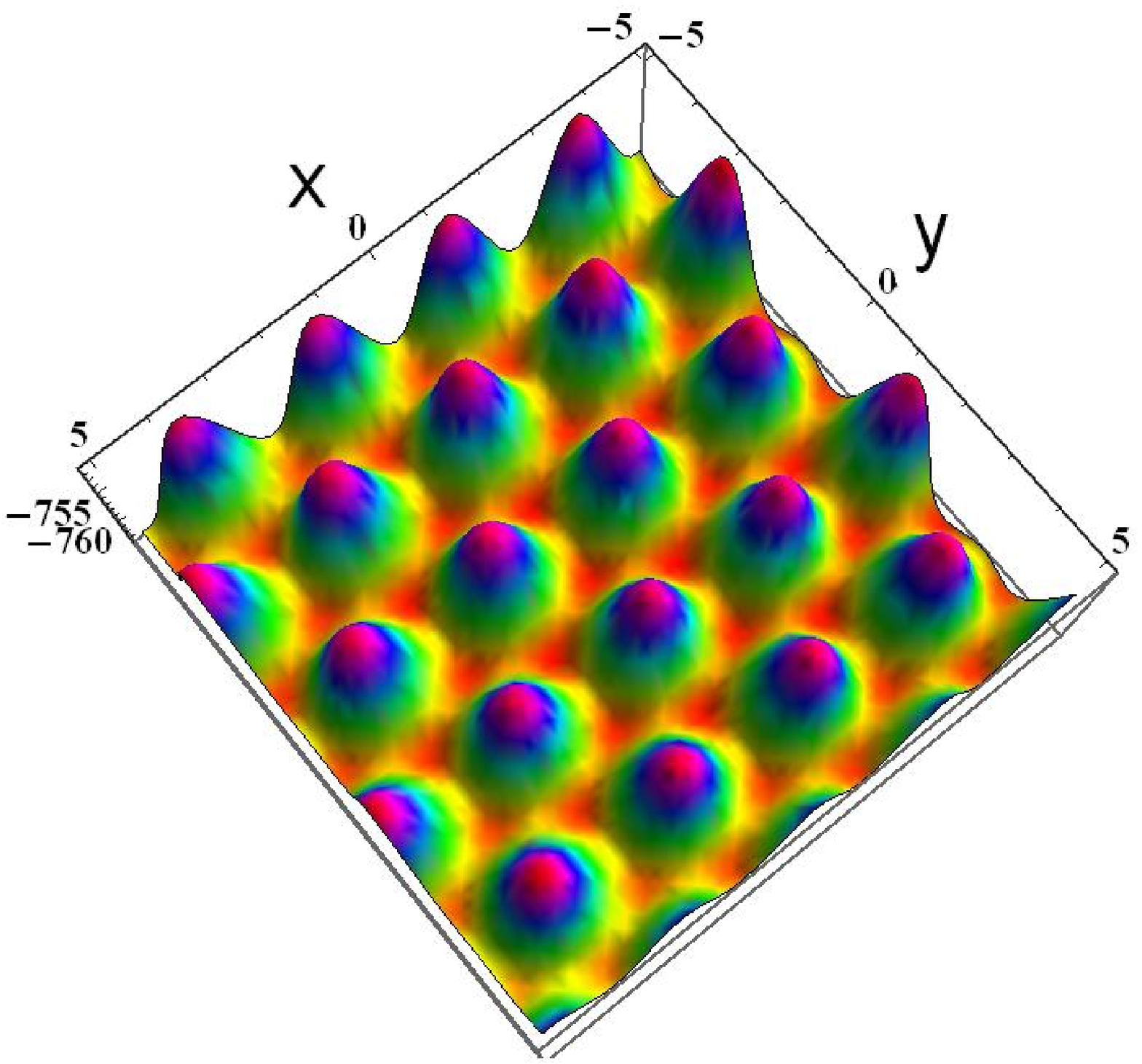}
\caption{(Color online) Periodic two dimensional potential created
by the graphene sheet. In this case $x$ and $y$ are the center of
mass coordinates
 of $C_{60}$ measured from an origin on the
sheet at the height $z=6.5$\AA. This is obtained by summing the
potential experienced by all individual atoms comprising the
$C_{60}$ molecule. The panel in the left shows the top view of the
potential, and the right panel shows the side view. The middle panel
shows the potential energy between $C_{60}$ molecule and flat
graphene sheet at $x$=0 as a function of y for several height
values.
 \label{fig2Dpc60}}
\end{center}
\end{figure}

\begin{figure}[ht]
\begin{center}
\includegraphics[width=0.6\linewidth]{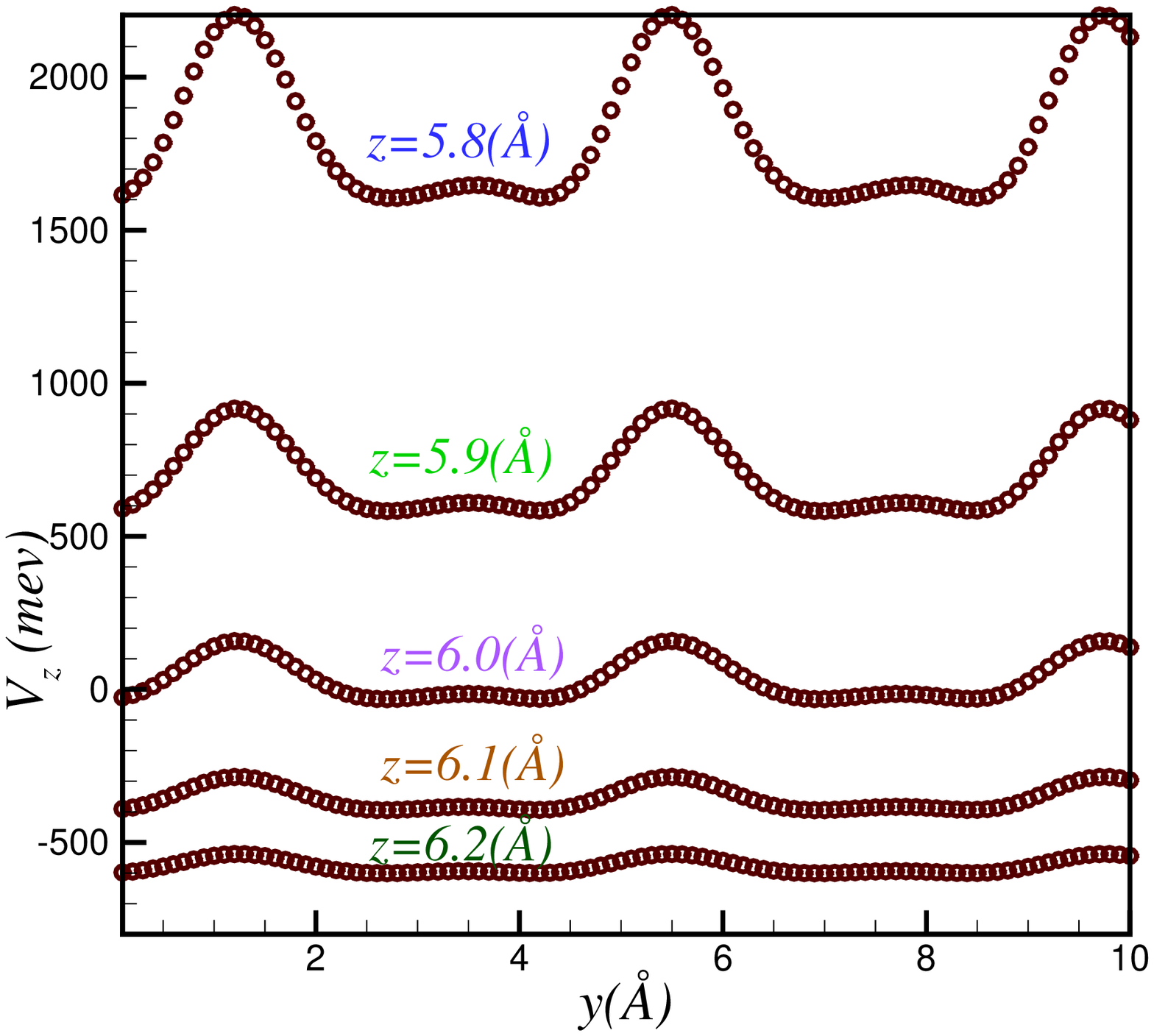}
\caption{(Color online) Periodic potential energy between $C_{60}$
molecule and flat grapheen sheet at $x$=0 as a function of y for
several height values.
 \label{vyc60}}
\end{center}
\end{figure}

\section{Two dimensional potential of garphene sheet }

First we study the periodic two dimensional LJ potential that a flat
graphene sheet creates near its surface. Generally,
  graphene is not a flat sheet in finite temperature and
  exhibits small roughness; it
may thus be locally approximated with a flat sheet \cite{nima}. The
aforementioned quantity helps one to gain some insight into how
graphene influences molecules such as $C_{60}$ moving at its
proximity.

A flat graphene sheet comprises two triangular Bravais sublattices
 (see Fig.~\ref{figAB}),  and the LJ potential due to
this sheet~\cite{Hendrick} can be written as the sum of these two
sublattice potentials
\begin{equation}\label{Eq1}
E_T(x,y,z)=V_{A-lattice}+V_{B-lattice},
\end{equation}
where $(x,y,z)$ is the position for center of mass of the molecule,
having $N$ atoms, above the sheet. We may write the appropriate
expression for $V_{A-lattice}$ as
\begin{equation}\label{VA}
V_{A-lattice} ={\sum_{m,n}}{\sum^N_{l=1}}{\sum^2_{k=1}}
\frac{4(-1)^{k+1}\sigma^{12/k}}{((x+x_{l}-n-m/2)^2+(y+y_{l}-\sqrt{3}m/2)^2+(z+z_{l})^2)^{6/k}}
\end{equation}
and for $V_{B-lattice}$ as
\begin{equation}\label{VB}
V_{B-lattice}={\sum_{m,n}}{\sum^N_{l=1}}{\sum^2_{k=1}}\frac{4(-1)^{k+1}\sigma^{12/k}}
{((x+x_{l}+1/2-n-m/2)^2+(y+y_{l}-\sqrt{3}m/2-\sqrt{3}/6)^2+(z+z_{l})^2)^{6/k}}
\end{equation}
Here $m$ and $n$ are integer numbers which count lattice points for
each sublattice. The coordinate of the $l^{th}$ atom
($x_{l},y_{l},z_{l}$) is measured from the molecule's center of
mass. The sum over $k$ also, is responsible for varying of
fraction's power between $12$ or $6$, and switching its sign.
Because of  the short range behavior of the LJ potential, we
observed that using the cut-offs of $|m|\geq 10$ or $|n|\geq 10$
leads to quite accurate results with negligible cut-off errors in
the total potential/force. For simplicity in Eqs.~(\ref{VA})
and~(\ref{VB}) all lengths were re-scaled by $a_1=\sqrt{3}a_0$ where
$a_0=1.42$~\AA~and energies are in units of $\varepsilon$, with
$a_1$ being the length of primitive vector of sublattices (see Fig.
\ref{figAB}).

\subsection{Potential energy between a single carbon atom and graphene sheet}

At a fixed height above the graphene sheet, $z=z_0$, the total
potential in Eq.~(\ref{Eq1}) reduces to a two-dimensional potential
$\tilde{V}(x,y)$, which is periodic in $x$ and $y$. Assume that we
have fixed the height of a point like particle,
 such as a carbon atom, at a given value.
 Figure~\ref{fig2Dp} shows this periodic two-dimensional potential
 for $z_0$=3.5~\AA. This potential provides some insights into the
 diffusive motion or trapping of a point-like particle close to the graphene sheet~\cite{nature2008}.
The required potential can be obtained from Eqs.~(\ref{VA}) and
(\ref{VB}) with $N=1$ and putting $x_{l},~y_{l}$ and $z_{l}$ equal
to zero. Figure~\ref{vy} shows the variation of potential energy in
$y$ direction at $x=0$ for five different heights. To show that the
variation of $\tilde{V}(x,y)$ in the $z $ direction appears to be similar to the LJ
potential, with different functionality, we averaged the above
potential in the Wigner-Seitz primitive cells on both sublattices
and plot the result in the bottom panel of Fig.~\ref{vz}. As can be
seen from the figure the potential minimum between graphene sheet
and the single carbon atom is deeper than the simple case of two interacting
carbon atoms via the LJ interaction. Furthermore, the minimum distance
($z_{min}\sim$3.4~\AA) is not $z=2^{1/6}\sigma\sim3.8$~\AA~ as it is
in the usual LJ (solid blue curve in Fig.~\ref{vz}). The minimum
point of $V_z$ gives the equilibrium distance of a carbon atom over
graphene flat sheet.

Note that the particle will equilibrate with the graphene sheet and
obtains a mean kinetic energy of the order of $K_{B}T$ in its 2D
lateral motion. If $K_{B}T \simeq 25.7$ meV is smaller than the
potential barriers heights, the particle may be trapped in one of
the potential wells. Since $\tilde{V}(x,y)$ is a two-dimensional
potential, the particle could take various paths from one well to
neighboring wells. But for simplicity we may roughly take the
difference between maximum and minimum values of $\tilde{V}(x,y)$
appearing in Wigner-Seitz primitive cells as a measure of the
barrier strength compared to the thermal energy $K_{B}T$. The top
panel in Fig.~\ref{vz} shows the difference between  maximum and
minimum $\Delta V$ in units of $K_{B}T$ for various values of the
particle height.

\subsection{Potential energy between $C_{60}$ molecule and graphene sheet}

\begin{figure}[ht]
\begin{center}
\includegraphics[width=0.6\linewidth]{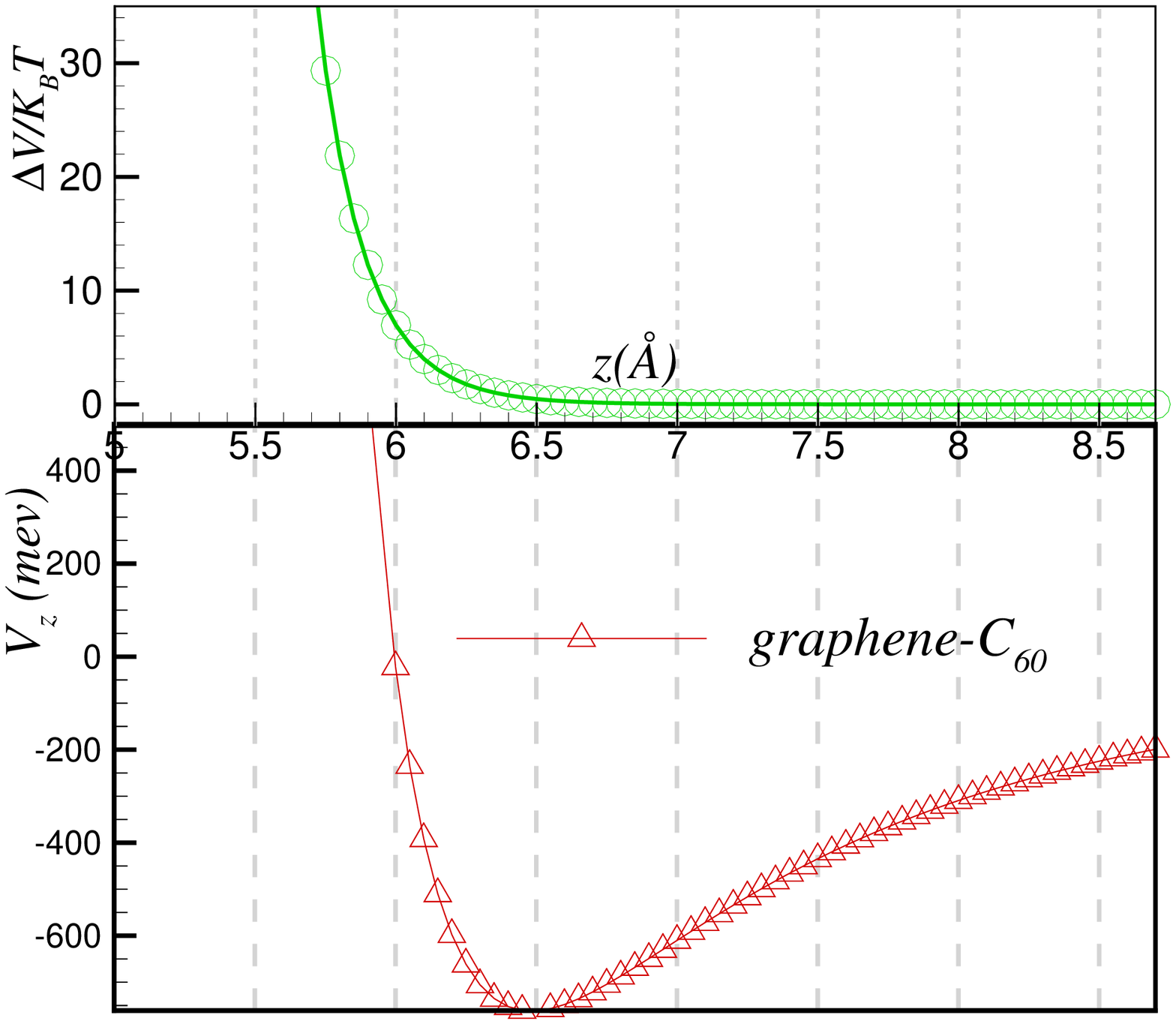}
 \caption{(Color online) Top: Difference between maximum and minimum of
two-dimensional potential  created by a flat graphene sheet (see
Fig. \ref{fig2Dpc60}) as a function of normal distance $z$ as
experienced by $C_{60}$ molecule . Bottom: Total potential energy
between a flat graphene sheet and a $C_{60}$ molecule. In both
panels, to eliminate the dependency of the $x$ and $y$ variables, we
averaged over Wigner-Seitz primitive cells of both A-lattice and
B-lattice with the data have infinitesimal ($10^{-5}$) error bars
which were not shown.\label{c60vz} }
\end{center}
\end{figure}

In the case of molecules such as $C_{60}$ as considered here, the
total two-dimensional potential is periodic as well, see
Fig.~\ref{fig2Dpc60}. The interaction between this molecule and the
graphene sheet has several functional forms depending on the
orientation of the molecule over the sheet even at a fixed height of
the molecule's center of mass above the sheet. Two particular
orientations of $C_{60}$ are more interesting than the others. These
two refer to the cases when a pentagon or a hexagon of $C_{60}$'s
atoms faces the flat graphene surface. Figure~\ref{fig2Dpc60} is
related to a $C_{60}$ molecule when one of the pentagons is  near to
the surface and the plane of the pentagon is parallel to the
graphene sheet. Since the radius of the $C_{60}$ molecule is
$R_{C_{60}}$=3.54~\AA, it can never get closer to the surface than
this distance. Figure~\ref{vyc60}~ shows the variation of potential
energy along normal direction for five different height of the
center of mass of $C_{60}$ molecule. Furthermore the variation of
$E_T$ along $z$ direction averaged over Wigner-Seitz primitive cells
of A-lattice and B-lattice is shown in bottom panel of
Fig.~\ref{c60vz}. This gives the minimum height value for the center
of mass of $C_{60}$ molecule as $z_{min}\sim~6.5$~\AA~which is
obviously larger than the equilibrium distance obtained for a single
carbon above the flat sheet. The binding energy of $C_{60}$ molecule
and monolayer graphene can be estimated around 800~mev which is the
same to the experimental value for binding energy of $C_{60}$
molecule and graphite bulk~\cite{Hendrick}.

Similar to the the single carbon case we show the difference between
maximum and minimum appearing in Wigner-Seitz primitive cells with
respect to the thermal energy in top panel of Fig.~\ref{c60vz}. The
several orientations have not much effects in the curves
in~Fig.~\ref{c60vz}. Furthermore, the height of saddle points in the
potential profile in Fig.~\ref{fig2Dpc60} with respect to the
thermal energy is around 0.2, which indicates that thermal energy is
the dominant factor. Therefore, we expect that there should not be
any trapping in the motion of $C_{60}$ over graphene even at low
temperatures and various orientations. In the next section we show
that the motion coincides with a diffusive Brownian motion.

This foregoing observation about the size of $C_{60}$ molecule and
the ratio between thermal energy and the depth of the potential
wells may be extended to other macromolecules and nanostructures
such as bucky-balls and carbon nanotubes. Almost all of these carbon
allotropes have longitudinal and lateral dimensions larger than
those of $C_{60}$. In a future work we will investigate dynamics of
other carbon allotropes over this mono layer sheet at finite
temperatures.

\section{Translational diffusion}

\begin{figure}[ht]
\begin{center}
\includegraphics[width=0.6\linewidth]{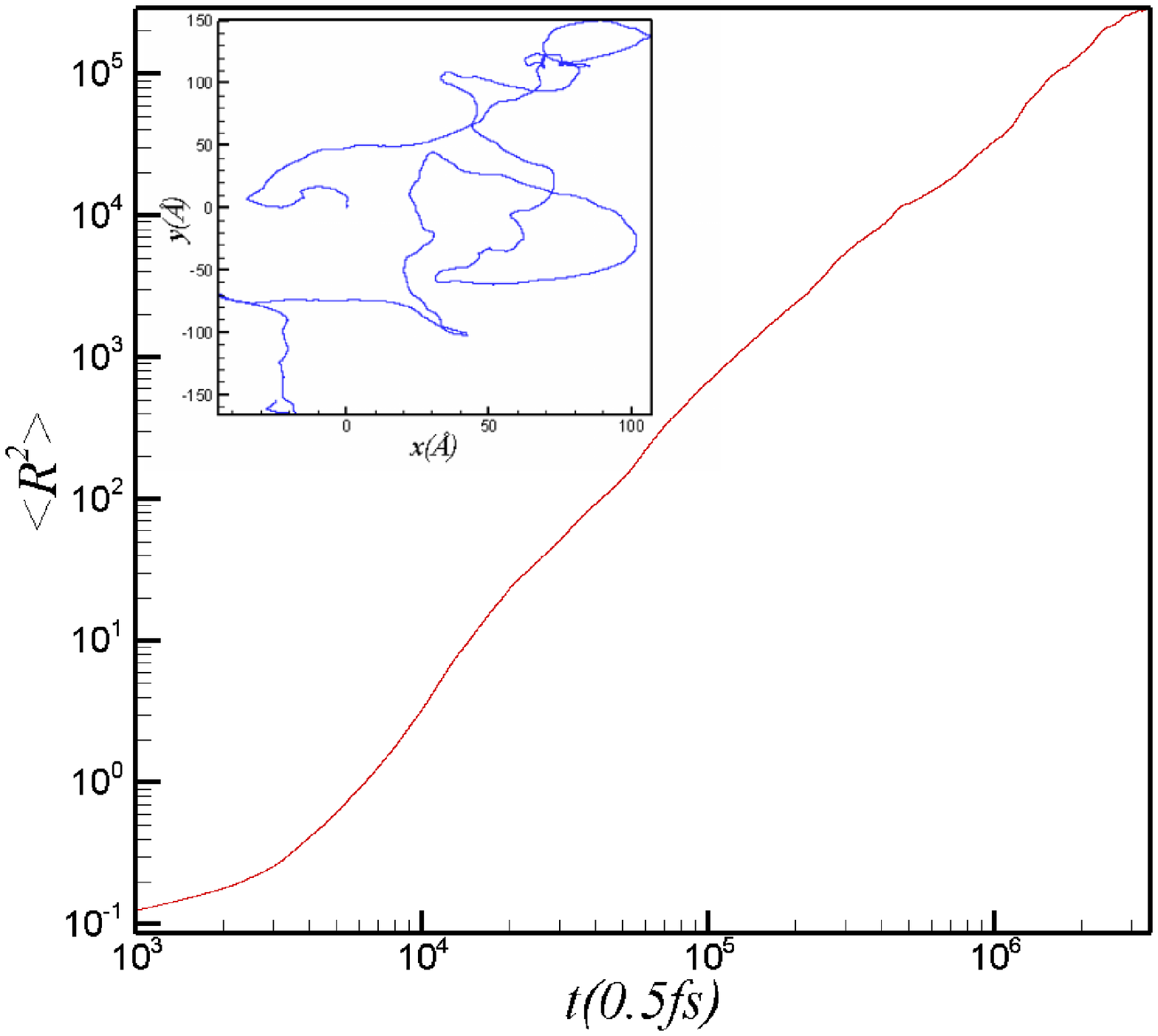}
\caption{(Color online) Mean square displacement, $\langle
R^2\rangle$, versus time for the lateral motion of an ensemble of 30
$C_{60}$ molecules moving near a graphene sheet
 at room temperature.
Inset show a typical $x-y$ trajectory of a single $C_{60}$ on the
graphene sheet. \label{snapshot} }
\end{center}
\end{figure}

In Fig. \ref{snapshot}, the mean square displacement $\langle
R^2\rangle$ of an ensemble of 30 $C_{60}$ molecules moving near a
graphene sheet at room temperature is plotted as a function of time
for the projected two dimensional motion onto the x-y plane, where
 $R^2=x^2+y^2$. The inset of the figure shows $x-y$ trajectory for a single
$C_{60}$ molecule. The total simulation time is 2.5~ns. Diffusion
coefficient for the $C_{60}$ molecule is obtained from this graph as
$D=7.0\times10^{-10}{\mathrm{m}}^2{\mathrm{s}}^{-1}$. Using
Einstein's relation, we estimate an effective friction coefficient
of
$\xi=\frac{K_BT}{D}=1.4\times10^{9}{\mathrm{m}}^{-2}{\mathrm{s}}\times
K_BT$ for
 $C_{60}$ moving near a graphene sheet.
 At very short times the observed motion is
not diffusive because we put all $C_{60}$ at the center of the
membrane with initial random velocities extracted from a
Maxwell-Boltzman distribution, and the graphene is not in thermal
equilibrium with those molecules. Therefore at very short time of
about 5~ps, the $C_{60}$ molecules attempt to find the minimum
energy trajectories, yet their total displacement is not
significant.

\section{Effective potential for vibrational motion in $z$-direction}

\begin{figure}[ht]
\begin{center}
\includegraphics[width=0.6\linewidth]{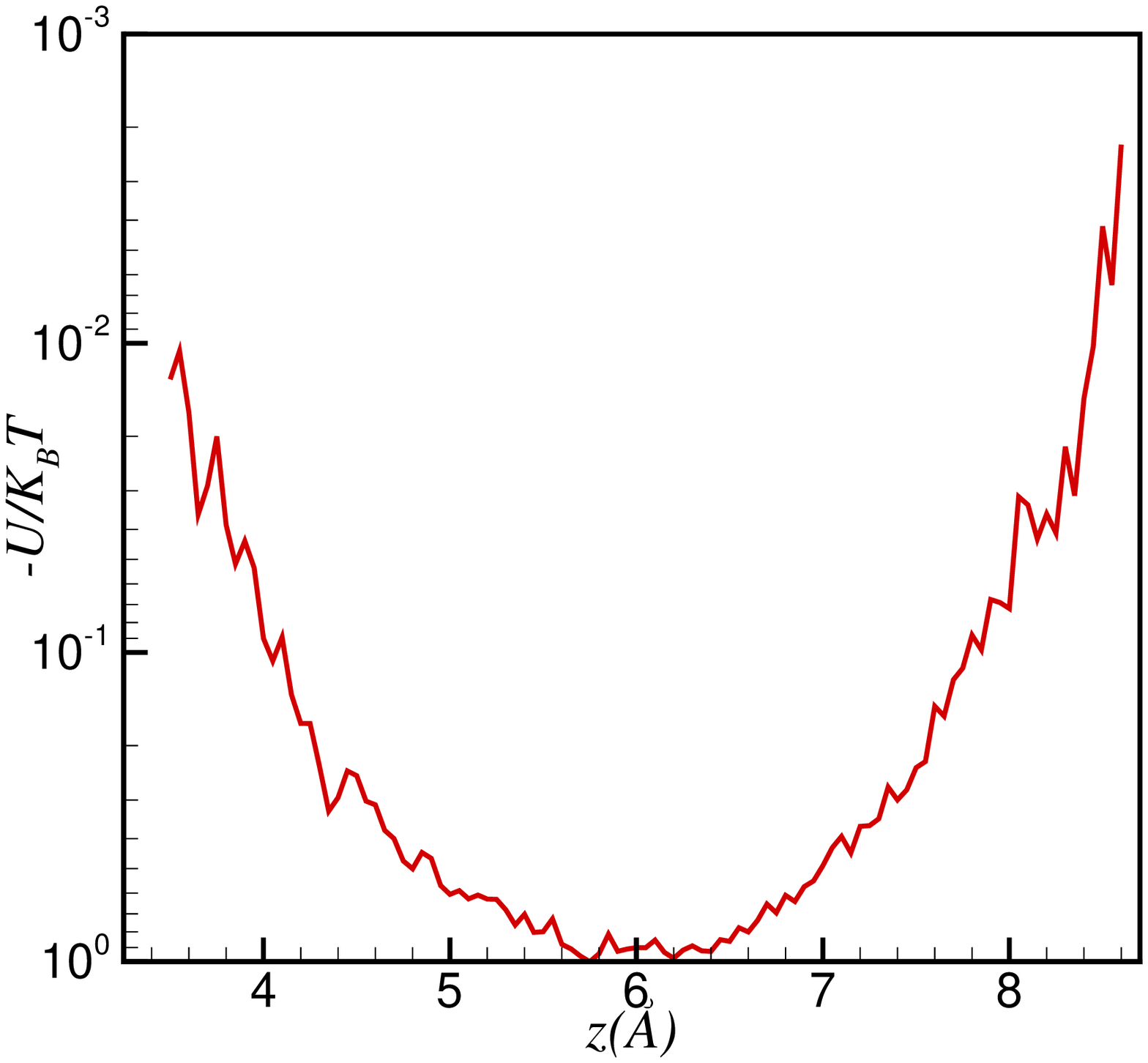}
\caption{(Color online) Effective potential of the interaction
between $C_{60}$ molecule and a graphene sheet. \label{U_pot}}
\end{center}
\end{figure}

In our MD simulations the $C_{60}$ molecule does not unbind from the
graphene sheet. We may introduce an effective potential for the
$C_{60}$ motion  in the $z$-direction by calculating the
distribution function, $p(z)$, of the height of a $C_{60}$ above the
sheet. By virtue of the Boltzmann factor, we may define the
effective potential as $\frac{U}{K_{B}T}=-\log[p(z)]$ which embeds
in itself both entropic and energetic factors related to the
equilibrated motion of $C_{60}$ near the graphene sheet.  Results
are represented in Fig.~\ref{U_pot}. It turns that the effective
potential has a harmonic-type shape and
 $\frac{U}{{K_{B}T}}\cong
\frac{1}{2}k (z-z_m)^2$ with the effective parameters $k$ and $z_0$
obtained from a best fit as
$k=1.255K_{B}T~({\mathrm{N}}/{\mathrm{\AA}})$ and $z_m=5.908$~\AA.
One thus expects that the $C_{60}$ molecules effectively exhibit
bounded vibrational motion in the normal direction to the graphene
sheet. In lateral directions, as we discussed in the previous
sections, the motion is diffusive. Here the mean value for $\langle
z\rangle$=5.975~\AA~ shows the equilibrium distance of center of
mass of $C_{60}$ molecule  which already has been thermalized with
the graphene sheet. This value of height depends on the temperature
of the system. Note that this equilibrium distance is calculated for
the rough graphene unlike the value reported in section 3. The
distance from graphene sheet is thus found to be bigger than the
distance of $C_{60}$ from the first layer of gold bulk
\cite{small2007}, i.e. 5.4~\AA. This is because the graphene sheet
is a single atomic layer instead of a bulk material.

\section{Rotational diffusion}

Rotation matrix  transforms the coordinates of a vector in
body-fixed frame to the coordinates in the lab-fixed frame. We
define the body-fixed frame as two perpendicular vectors, which are
chosen as two lines connecting two pairs of opposite points on the
$C_{60}$ cage and the cross vector of those two vectors. In order to
be sure about the orthonormality of those vectors the Gram-Schmidt
orthonormalization processes was applied in each time step of
simulation.

The quaternion representation of rotation matrix is a useful method
for numerical simulations since they are more numerically stable.
~\cite{md}.
 The angular velocities of $C_{60}$ in the body-fixed frame can be written in the term of
 quaternions and their time derivatives.
Applying the inverse of the rotation matrix on the  angular velocity
of $C_{60}$ in the body-fixed frame, yields the angular velocities
of $C_{60}$ in lab-fixed frame, $\overrightarrow{\omega}$.


To show that motion of $C_{60}$ on the graphene is not a rolling
motion, we have calculated the cross-correlation of the unit vector
of angular velocity, $\widehat{\omega}$, and the unit vector of velocity,
$\widehat{v}$, i.e, $\langle \widehat{\omega}\cdot\widehat{v}
\rangle =3.0\times10^{-2}\pm 0.01$ and $\langle
|\widehat{\omega}\cdot\widehat{v}| \rangle =0.47\pm 0.01$. Note that the first
average is almost zero but it can not confirm that these two vectors are
uncorrelated because this average
becomes zero  for two perpendicular vectors as well. On the other hand, the non-zero second average shows
that they are not perpendicular. Therefore, these two averages
together make sure that these two vectors are uncorrelated. The
independence of the direction of velocity from the angular velocity
ensures that the motion of $C_{60}$ over graphene sheet is not a
rolling motion.

\begin{figure}[ht]
\begin{center}
\includegraphics[width=0.6\linewidth]{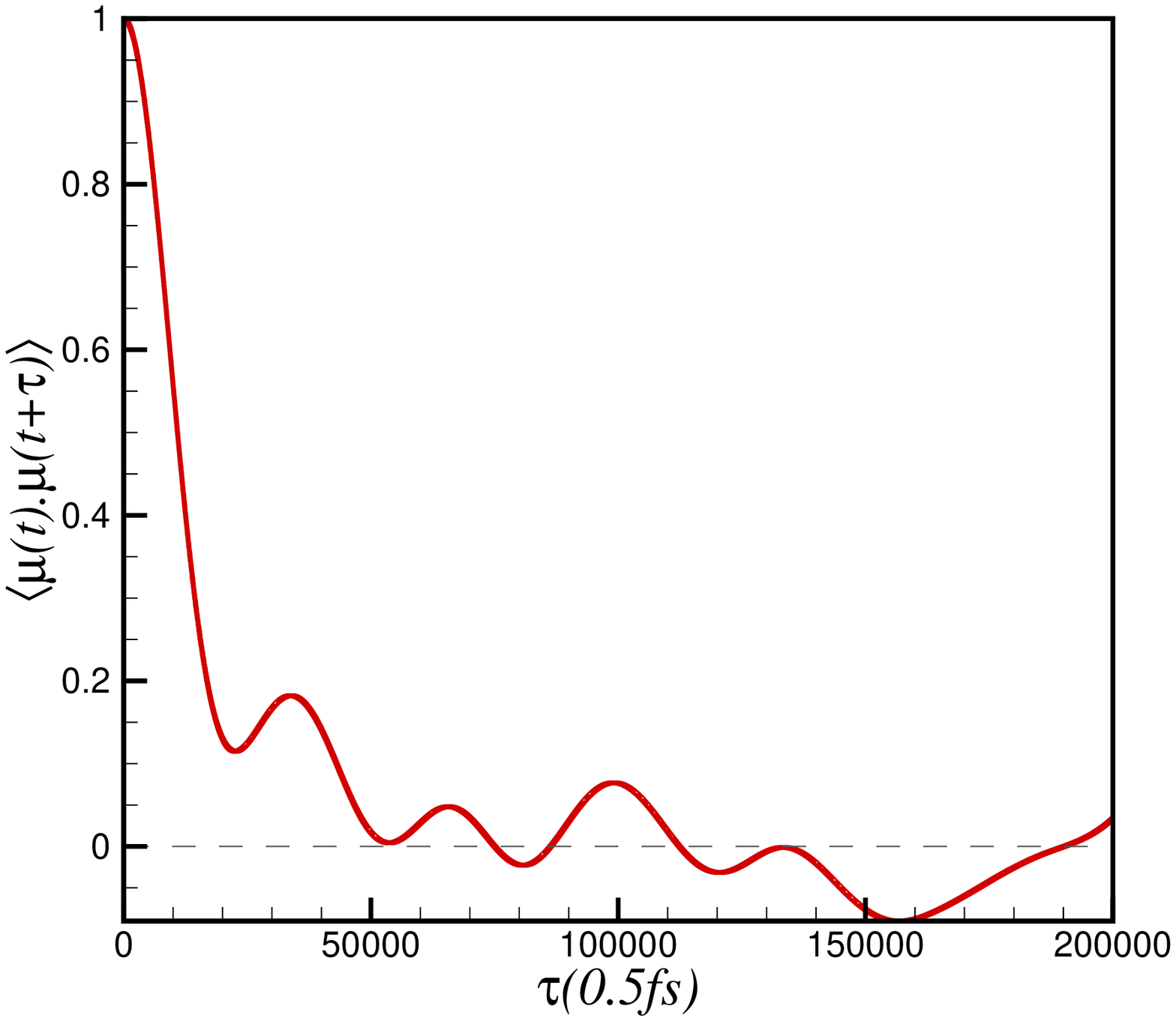}
\caption{Rotational correlation: the correlation length of $\mu$ is
about 75000 steps (37.5~ps). \label{miyou}}
\end{center}
\end{figure}

\begin{figure}[ht]
\begin{center}
\includegraphics[width=0.6\linewidth]{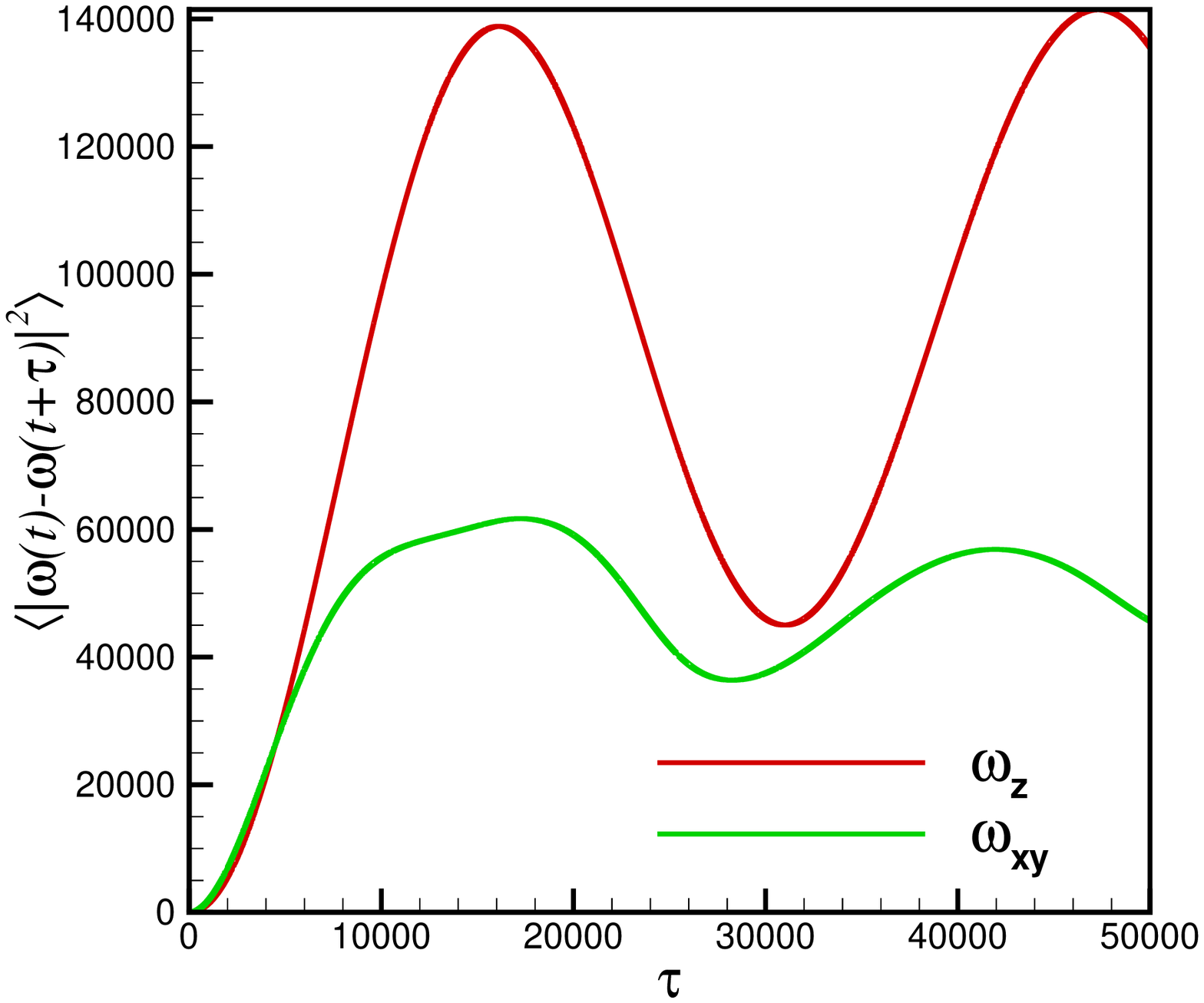}
\caption{(Color online) Increment of $\omega_z$ and $\omega_{xy}$
versus time step. \label{wzxy}}
\end{center}
\end{figure}

Defining a fixed vector in $C_{60}$, $\mu$, helps us to investigate
the diffusive nature of $C_{60}$ motion . It can be understood from
the time auto correlation of $\mu$  where the correlation time is
$37.5$~ps (see Fig. \ref{miyou}). It is obvious that the correlation
length of $\mu$ is about 75000 steps  which means that after this
time the orientation of $C_{60}$ becomes completely different.
Angular velocity in different directions has different behaviors.
Figure \ref{wzxy} shows the increment of $\omega_z$ and
$\omega_{xy}$. Obviously the fluctuations of $\omega_z$ is bigger
than $\omega_{xy}$. This is due to the structure of $C_{60}$ which
is not a continues ball. It is a discrete spherical object. As we
mentioned in the previous section the pairwise interaction of
$C_{60}$ with the graphene atoms depends on the several orientations
of $C_{60}$. For example when a hexagon of $C_{60}$ is parallel to
another hexagonal in the graphene plane, $C_{60}$ would be more
stable than the other possible orientations. These restrictions do
not affect the rotation around the $z$ axis,  so the rotation around
the $z$ axis is easier than around $x-y$ direction.

\section{Conclusion}
We studied the motion of $C_{60}$ molecule over a graphene sheet. Both
flat approximation for monolayer graphene sheet and monolayer
graphene at a finite temperature have been studied using atomistic
simulations. The depth of the potential wells generated by a graphene sheet
in its proximity decreases as the height
of an external object increases above the sheet. The binding energy of
$C_{60}$ over a graphene sheet was found as 800~meV close to the
experimental value for binding energy of $C_{60}$ and
graphite~\cite{Hendrick}. The motion of $C_{60}$ in the
perpendicular direction was found to be a vibrational motion similar to
a simple harmonic oscillator. While the motion in lateral
directions  is found to be a diffusive non-rolling motion.

\pagebreak
\end{document}